\documentstyle[multicol,aps,epsf]{revtex}
\begin{document}
\draft

\title{A Distribution of Tunnel Splittings in Mn$_{12}$-Acetate.}
\author{K. M. Mertes, Yoko Suzuki, and M. P. Sarachik}
\address{Physics Department, City College of the City University of New York,
New York, NY 10031}
\author{Y. Paltiel, H. Shtrikman, and E. Zeldov}
\address{Department of Condensed Mater Physics, 
The Weizmann Institute of Science, Rehovot 76100, Israel}
\author{E. Rumberger and D. N. Hendrickson}
\address{Department of Chemistry and
Biochemistry, University of California at San Diego, La Jolla, CA 92093}
\author{G. Christou}
\address{Department of Chemistry, Indiana University, Bloomington, Indiana
47405}
\date{\today}
\maketitle 

\begin{abstract}
In magnetic fields applied parallel to the anisotropy axis, the relaxation of the
magnetization of a Mn$_{12}$-acetate single crystal measured for different sweep rates is shown to collapse
onto a single scaled curve.  The form of the scaling implies that the dominant
symmetry-breaking process that gives rise to tunneling is a locally varying
second-order anisotropy, forbidden by tetragonal symmetry in the perfect crystal,
which gives rise to a broad distribution of tunnel splittings in a real crystal of
Mn$_{12}$-acetate.  Different forms applied to even and odd-numbered steps provide
a clear distinction between even step resonances (associated
with crystal anisotropy) and odd resonances (which require a transverse component of
magnetic field).
\end{abstract}

\begin{multicols}{2}
Single-molecule magnets are organic materials which contain a
large (Avogadro's)  number of identical magnetic molecules; 
([Mn$_{12}$O$_{12}$(CH$_3$COO)$_{16}$(H$_2$O)$_4$]$\cdot$ 2CH$_3$COOH$\cdot4$H$_2$O), 
generally referred to as Mn$_{12}$-acetate, is a particularly interesting and
much-studied example of this class.  The Mn$_{12}$ clusters are composed of twelve
Mn atoms tightly coupled to give a sizable $S=10$ spin magnetic moment that is stable
at temperatures of the order of $10$ K and below\cite{sessoli}.  These identical
weakly-interacting magnetic molecules are regularly arranged on a tetragonal crystal.

\begin{figure}
\centerline{\epsfxsize 0.8\columnwidth
\epsfbox{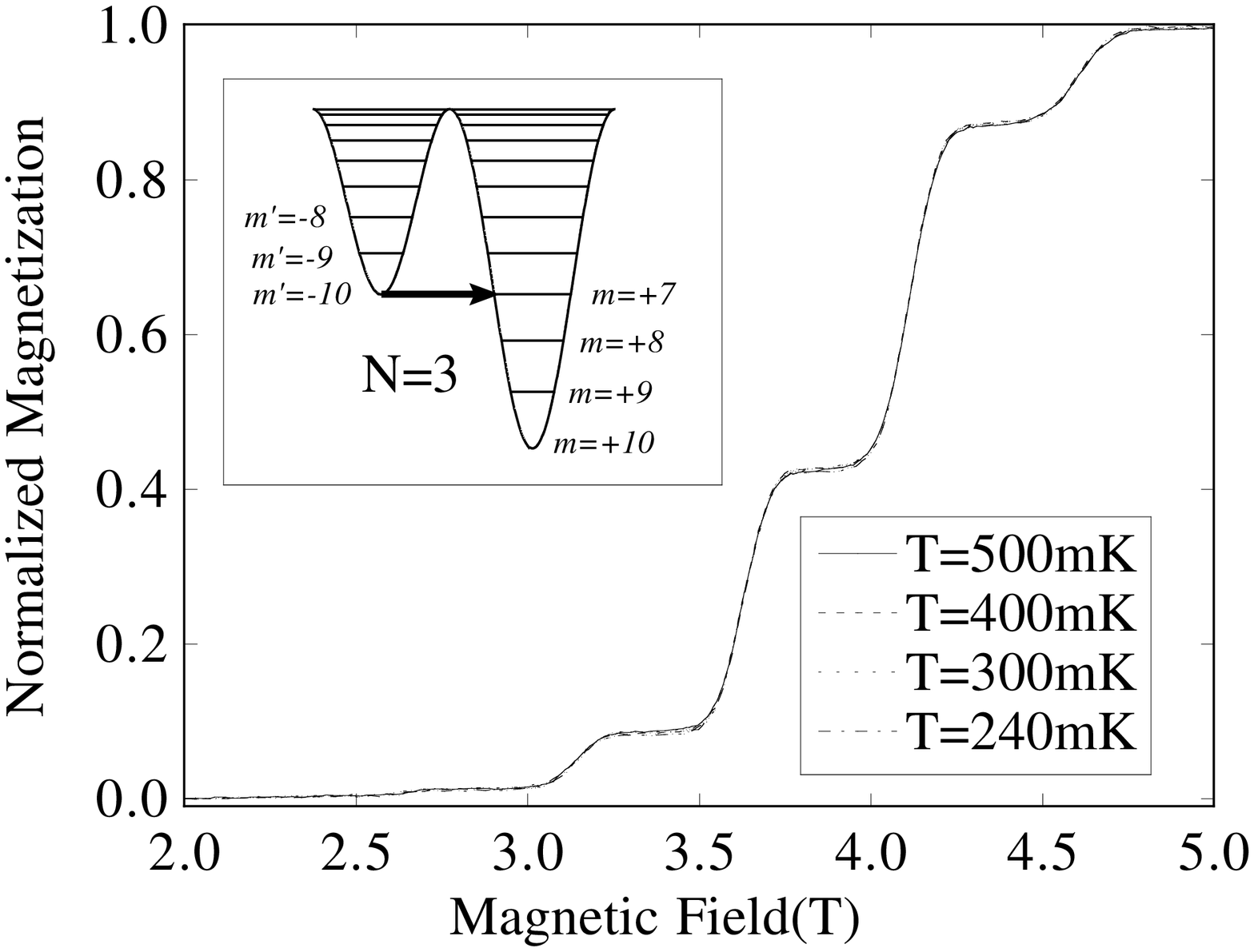}}
\caption{Magnetization versus longitudinal magnetic field for a
Mn$_{12}$ sample starting from a demagnetized state;
data are shown for the same sweep rate at different temperatures ranging from
$0.500$K to $0.240$K.  The curves overlap indicating that
relaxation is from the ground state, as shown for the double-well potential
illustrated in the inset.}
\label{TempIndependence}
\end{figure}

As illustrated by the double well potential shown in the inset to Fig.
\ref{TempIndependence}, strong uniaxial anisotropy (of the order of $65$
K) yields a set of energy levels corresponding to different projections
$m = \pm 10, \pm 9,.....,0$ of the total spin along the easy
c-axis of the crystal.  Measurements\cite{friedman,thomas} 
below the blocking temperature of $3$ K have revealed a series
of steep steps in the curves of $M$ versus $H$ at  roughly
equal intervals of magnetic field due to enhanced relaxation 
of the magnetization whenever levels on opposite sides of the
anisotropy barrier coincide in energy. Below $\approx 0.56$K the
magnetization curves are independent of temperature, and the tunneling
proceeds from the ground state of the metastable well (see 
inset to Fig. \ref{TempIndependence}).  

The spin Hamiltonian for Mn$_{12}$ is given by:
\begin{equation}
\label{eq_hamiltonian}
{\cal H} = -D S_z^2 -g_z \mu_B^{} H_z S_z - A S_z^4 + .......
\end{equation}
where $D=0.65$ K is the longitudinal anisotropy, the second term is the
Zeeman energy with $g_z\approx1.94$, and the third on the right-hand side represents
the next higher-order term in longitudinal anisotropy.  In order for tunneling to
occur, the Hamiltonian must also include terms that do not commute with
$S_z$.  In a perfect crystal, the lowest transverse anisotropy term allowed
by the tetragonal symmetry of Mn$_{12}$ is proportional to $(S_{+}^4 +
S_{-}^4)$.  For ground state tunneling, such a term only permits every
fourth step.  In contrast, all steps are observed with no clear differences
in amplitude between them.  This suggests that transverse internal magnetic
fields, which would allow all steps to occur on an equal footing, provide
the dominant symmetry-breaking term that drives the tunneling in Mn$_{12}$. 
However, dipolar fields \cite{prokofev,chud1,prokofev2,wernsdorfer} and
hyperfine interactions \cite{prokofev,boutron,chud2} are too weak to cause
the rapid tunneling rates observed; the nature of the effects responsible
for tunneling in Mn$_{12}$ has remained an open question.

In this paper we report data obtained for the relaxation of the magnetization of
Mn$_{12}$ in a swept field for different sweep rates.  We show that a scaling
form recently proposed by Garanin and Chudnovsky
\cite{dislocation1,dislocation2}, who considered the effect of crystal
dislocations, yields an approximate collapse of all the data onto a single
curve.  The form of the scaling function corresponds to tunneling due to
second-order transverse anisotropy that varies throughout the Mn$_{12}$ crystal
with a very broad distribution.  Departure from perfect scaling is
observed that is associated with a small admixture of tunneling due to other
symmetry-breaking terms, presumably transverse internal magnetic fields.  
Thus
\begin{equation}
\label{eq_hamiltonian_complete}
{\cal H} = ..... + E(S_x^2-S_y^2) - H_x S_x.
\end{equation}
with $E=E(x,y,z)$ and $H_x=H_x(x,y,z)$ varying from point to point in the Mn$_{12}$
crystal.  Our results imply that the dominant term responsible for tunneling
is second order anisotropy which, although prohibited in a perfect
crystal, is present and significant in real crystals of Mn$_{12}$.

The magnetization of small single crystals of Mn$_{12}$-acetate
was determined from measurements of the local magnetic
induction at the sample surface using $10 \times 10$
$\mu$m$^2$  Hall sensors composed of a two-dimensional electron
gas (2DEG) in a GaAs/AlGaAs  heterostructure.  The 2DEG was
aligned parallel to the external magnetic field, and the Hall
sensor was used to detect the perpendicular component (only) of
the magnetic field arising from the sample magnetization
\cite{zeldov}.

\begin{figure}
\centerline{\epsfxsize 1\columnwidth
\epsfbox{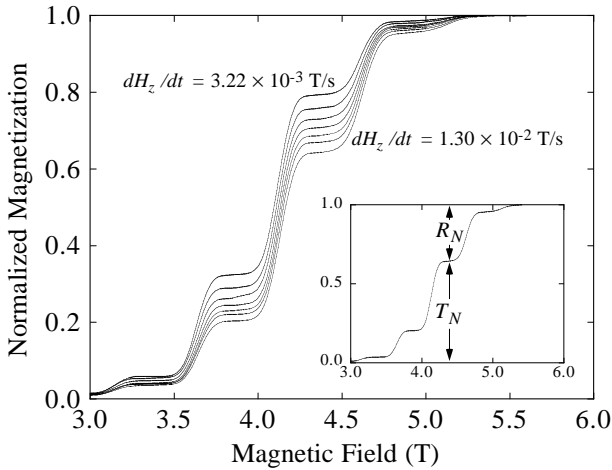}}
\caption{Normalized magnetization curves at $0.24$ K for sweep rates ranging
from $1.30 \times 10^{-2}$ T/s to $3.22 \times 10^{-3}$ T/s.  The plateaux indicate the
cumulative fraction of molecules that have tunneled after an energy level crossing.  For clarity only a
partial data set is presented here.  The inset illustrates the definitions of $T_N$ and
$R_N$ discussed in the text.}
\label{SweepRateDependence}
\end{figure}

\begin{figure}
\centerline{\epsfxsize 1\columnwidth
\epsfbox{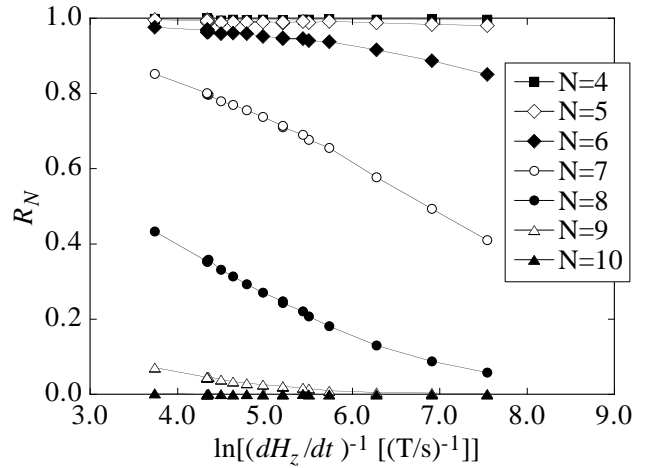}}
\caption{The fraction of molecules, $R_N$, that remain in the
metastable well following the $N^{th}$ level crossing plotted
as a function of the logarithm of $( d H_z /d t)^{-1}$.}
\label{UnscaledRelaxation}
\end{figure}

The magnetization of a Mn$_{12}$-acetate crystal normalized by the saturation value is shown in Fig.
\ref{SweepRateDependence} for different sweep rates from $1.30 \times 10^{-2}$ T/s to $5.28 \times 10^{-4}$
T/s.  Each curve was obtained starting from a demagnetized state by cooling the sample in zero magnetic
field from above the blocking temperature to a low temperature where pure ground state tunneling is
observed.

Since the magnetization has been normalized by its value at saturation, the
data shown in Fig. \ref{SweepRateDependence} represents the $cumulative$ fraction of
molecules that have tunneled from the metastable into the stable well.  The normalized
magnetization at a plateau, labeled $T_N$ in the inset of Fig.
\ref{SweepRateDependence}, thus represents the cumulative fraction of molecules that
have tunneled (relaxed) following an energy level crossing.  By the same token, $R_N=1-T_N$ is the
fraction of molecules that remain in the metastable well after the $N^{th}$ level crossing. 
Fig. \ref{UnscaledRelaxation} shows $R_N$ versus the logarithm of the inverse field sweep
rate, $(d H_z /d t)^{-1}$.

For a single molecule, the probability of remaining in the metastable well is given by
the Landau-Zener formula $P_N=\exp(-\pi\Delta^2_N/2v_N^{})$
\cite{wernsdorfer,dislocation1,dislocation2,Landau,Zener,Miyashita,dobrovitsky,leuenberger},
where
$\Delta_N$ is the level splitting of the $N^{th}$ resonance and $v_N^{}$
is the energy sweep rate defined by
$v_N^{}=(g_z\mu_B^{}\hbar/k_B^2)(2S-N) d H_z /d t$.  To date, all
attempts to account for the measured probabilities using the Landau-Zener
method have required unreasonably large tunnel splittings.  As we show
below, a distribution of tunnel splittings,
$\Delta_{N,i}$ (where $i$ denotes the $i$th molecule), provides a more accurate
description of tunneling in real crystals of Mn$_{12}$-acetate.  

For a distribution of tunnel splittings, $\Delta_{N,i}$, the probability that a spin
remain in the metastable well must be averaged over all the molecules:
$\langle{P_{N,i}}\rangle=\frac{1}{N_{T}}\sum\limits_{i}\exp
(-\pi\Delta^2_{N,i}/2v_N^{})$, where $N_{T}$ is the total number of molecules.  If
the distribution is very broad, then $\langle{P_{N,i}}\rangle$ is best examined on a
log scale, where an exponential looks like a step function, so that 
$\exp(-\pi\Delta^2_{N,i}/2v_N^{})\approx\Theta(1-\pi\Delta^2_{N,i}/2v_N^{})$.  This means that
for a fixed field sweep rate, $( d H_z /d t)$, those molecules that have tunnel splittings
obeying:
\begin{equation}
\label{eq_deltaCondition}
\pi\Delta^2_{N,i} < 2v_N^{}
\end{equation}
will remain in the metastable well for each $N$.  In essence,
\begin{equation}
\label{eq_RN}
R_N=\langle{P_{N,i}}\rangle=\frac{1}{N_{T}}\sum\limits_{i}\Theta(1-\pi\Delta^2_{N,i}/2v_N^{}).
\end{equation}  
and each curve in Fig. \ref{UnscaledRelaxation} denotes
the $fraction$ of molecules that remain in the metastable well after each
energy resonance because they tunnel too slowly \cite{shift}.  Since $\Delta_{N,i}$ is greater than
$\Delta_{N-1,i}$, each step probes a set of molecules in the crystal belonging to a different
range of the (initial) distribution of tunnel splittings.

Eq. \ref{eq_RN} shows that varying the sweep rate changes the fraction of molecules that remain
in the metastable well after the field sweeps through a particular energy resonance.  In fact, $R_N$
is the fraction of molecules which satisfy Eq. \ref{eq_deltaCondition} up to the threshold condition
$\pi\Delta^2_{N,i} = 2v_N^{}$.  This suggests that varying the sweep rate provides a method for probing the
distribution of tunnel splittings.  

Chudnovsky and Garanin \cite{dislocation1,dislocation2} have recently considered a distribution of tunnel
splitting due to transverse anisotropies caused by crystal dislocations.  The formalism
developed in references
\cite{dislocation1,dislocation2} is applicable regardless of the physical origin of the distribution,
provided it is logarithmically wide.  Neglecting the quartic longitudinal anisotropy term $AS_z^4$, the
tunnel splitting of molecule $i$ due primarily to second-order transverse anisotropy is given by:
\begin{equation}
\label{eq_delta}
\Delta_{N,i}=\eta_N^{}g_N^{}\left(\frac{\mid{E_i}\mid}{2D}\right)^{\xi_N^{}},
\end{equation}
where $g_N^{}=\frac{2D}{[(2S-N-2)!!]^2}\sqrt{\frac{(2S-N)!(2S)!}{(N)!}}$.  Transverse anisotropy
generates selection rules that only allow even-numbered resonances.  The leading contribution to
even-numbered resonances involves $\xi_N^{}={S-N/2}$ virtual transitions with $\Delta m=\pm 2$; here
$\eta_N^{}=1$.  However, odd numbered resonances could also occur due to the
presence of internal transverse magnetic fields of hyperfine or dipolar origin, or transverse fields due
to the defects suggested by Chudnovsky and Garanin which locally tilt the easy axis so that the applied
longitudinal field has a transverse component\cite{dislocation2}.  The largest contribution to
odd-numbered resonances entails
$\xi_N^{}=S-(N-1)/2$ virtual transitions with $\Delta m=\pm 2$ due to the transverse anisotropy and a
single virtual transition with $\Delta m=\pm 1$ due to the transverse field.  One therefore expects that
the tunnel splitting for even and odd $N$ should be roughly comparable, as is observed.  For odd $N$, the
tunnel splitting has the same form as Eq. \ref{eq_delta}.  However, in this case, $\eta_N^{}=CN/2$ and
$\xi_N^{}=S-(N-1)/2$, where $C$ is an adjustable parameter of order $1$.

\begin{figure}
\centerline{\epsfxsize 1\columnwidth
\epsfbox{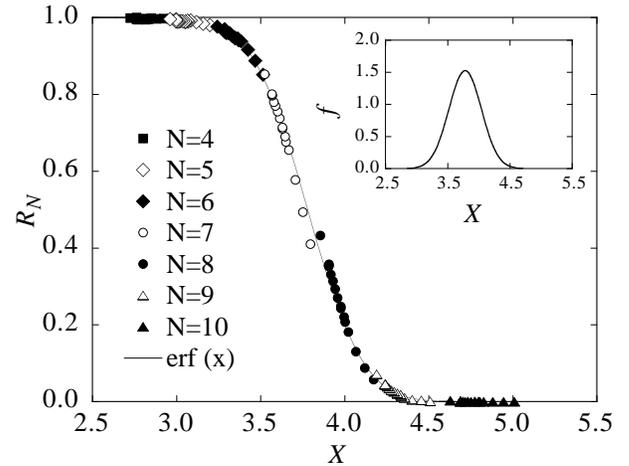}}
\caption{The fraction of molecules, $R_N$, that remain in the
metastable well following the $N^{th}$ level crossing plotted as a function of
the scaling parameter $X$.  The solid continuous curve is a best fit
to the data using the error function.  The inset illustrates the
Gaussian distribution of transverse anisotropies determined by taking the
derivative of this curve.}
\label{ScaledRelaxation}
\end{figure}

Rearranging Eq. \ref{eq_delta} and employing the threshold condition, $\pi\Delta^2_{N,i} = 2v_N^{}$, we
find that
\begin{equation}
-\ln\left(\mid{E_i}\mid/2D\right)=-\ln\left[\frac{1}{\eta_N^{}g_N^{}}\sqrt{\frac{2}{\pi}v_N^{}}\,\right]/\xi_N^{}\equiv
X
\end{equation}
is independent of $N$.  This implies that all the curves for
$R_N$ shown in Fig. \ref{UnscaledRelaxation} should collapse onto a single curve when plotted
as a function of $X$ (if second-order anisotropy is responsible for the tunneling).
This is shown in Fig. \ref{ScaledRelaxation}.  

We point out that scaling requires that different forms be used for even and odd-numbered steps,
providing a clear distinction between even resonances (associated
with crystal anisotropy) and odd resonances (which require a transverse component of
magnetic field).  For the odd resonance scaling a good fit was obtained for $C=1$.

Although the scaling obtained is of good quality, deviations should and do
occur.  This is due to the fact that the scaling function $X$ was calculated
exactly for even-numbered steps involving transverse anisotropy, while the expression
for $X$ is only approximate in the case of odd-numbered steps requiring an admixture
of effects due to transverse magnetic fields.  It should be noted that whenever more
than one process contributes, perfect scaling should not occur.

\begin{figure}
\centerline{\epsfxsize 1\columnwidth
\epsfbox{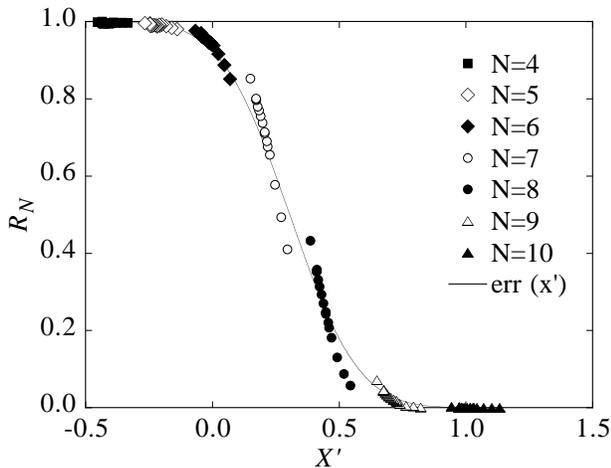}}
\caption{The scaling under the assumption that the tunnel splitting is due to
transverse field alone.}
\label{RelaxationTransverseField}
\end{figure}

The same formalism can be applied for the case of a distribution of tunnel splittings
due instead to a transverse field that varies throughout the sample.  In this case
the tunnel splitting has the form:
\begin{equation}
\label{eq_delta2}
\Delta'_{i,N}=g'_N\left(\frac{H_{x,i}}{2D}\right)^{(2S-N)},
\end{equation}
where $g'_N=\frac{2D}{[(2S-N-1)!]^2}\sqrt{\frac{(2S-N)!(2S)!}{(N)!}}$.  If the
tunnel splitting were due to transverse field alone the data should scale when
plotted as a function of the scaling parameter,
$-\ln\left(H_{x,i}/2D\right)=-\ln\left(\Delta'_{i,N}/g'_N\right)/(2S-N)\equiv X'$.  As can
be seen in Fig. \ref{RelaxationTransverseField}, the data do not scale by this procedure.  We
conclude that tunneling in Mn$_{12}$-acetate is primarily due to transverse anisotropy, as
evidenced by the approximate scaling shown in Fig. \ref{ScaledRelaxation}, with a
small admixture of tunneling due to transverse field which allows the odd
numbered resonances and gives rise to deviations from perfect scaling in
Fig. \ref{ScaledRelaxation}.

Since $T_N$ represents the cumulative fraction of molecules that have tunneled after the $N^{th}$ and all
previous crossings, the negative derivative of $R_N$ with respect to $X$ represents the distribution, $f$,
of transverse anisotropy.  The collapse of $R_N$ onto one curve supports the assumption that, rather than
being the same throughout the crystal as is usually assumed, the tunnel splittings vary locally within the
Mn$_{12}$ crystal with a very broad distribution.  The fact that the universal relaxation curve follows
the error function as shown by the solid continuous curve in Fig. \ref{ScaledRelaxation} indicates that the
distribution is approximately Gaussian\cite{dislocation2}.  The distribution is
shown in the inset of Fig. \ref{ScaledRelaxation}.

To summarize, the relaxation of the magnetization of Mn$_{12}$ measured in
longitudinal magnetic fields at different sweep rates collapses onto a
single scaled curve.  The form of the scaling implies that the dominant
symmetry-breaking process that gives rise to tunneling is a locally varying
second-order anisotropy, forbidden by tetragonal symmetry in the perfect
Mn$_{12}$-acetate crystal, which gives rise to a broad distribution of tunneling
splittings.  Different forms applied to even and odd-numbered
steps provide the first clear observation of a distinction between even step
resonances (associated with crystal anisotropy) and odd resonances (which require a
transverse component of magnetic field).

We thank D. Garanin and E. Chudnovsky for the numerous discussions that made this
analysis possible, and J. R. Friedman for valuable comments on the manuscript.  Work at City College was
supported by NSF grant DMR-9704309 and at the  University of California, San Diego by NSF grant
DMR-9729339.  EZ acknowledges the support of the German-Israeli Foundation for Scientific Research 
and Development.

\end{multicols}

\vspace{-.1in}
\begin{references}

\bibitem{sessoli} R. Sessoli, D. Gatteschi, A. Caneschi, and M. A. Novak, 
Nature {\bf 365}, 141-143 (1993).

\bibitem{friedman} J. R. Friedman, M. P. Sarachik, J. Tejada, and R. Ziolo,
Phys. Rev. Lett. {\bf 76}, 3830 (1996).

\bibitem{thomas} J. M. Hernandez, X. X. Zhang, F. Luis, J. Bartolome, J. Tejada, and
R. Ziolo, Europhys. Lett. {\bf 35}, 301 (1996); L. Thomas, F. Lionti, R. Ballou, R.
Sessoli, D. Gatteschi, and B. Barbara, Nature (London) {\bf 383}, 145 (1996).

\bibitem{prokofev} N. V. Prokof'ev and P. C. E. Stamp, Phys. Rev. Lett. {\bf 80},
5794 (1998).

\bibitem{chud1} E. M. Chudnovsky, Phys. Rev. Lett. {\bf 84}, 5676 (2000).

\bibitem{prokofev2} N. V. Prokof'ev and P. C. E. Stamp, Phys. Rev. Lett. {\bf 84},
5677 (2000).

\bibitem{wernsdorfer}  W. Wernsdorfer, C. Paulsen, and R. Sessoli, Phys. Rev. Lett.
{\bf 84}, 5678 (2000).

\bibitem{boutron} F. Hartmann-Boutron, P. Politi and J. Villain, Int. J. Mod. Phys.
B {\bf 10}, 2577 (1996).

\bibitem{chud2} D. A. Garanin, E. M. Chudnovsky, and R. Schilling, Phys. Rev. B {\bf
61}, 12 204 (2000).

\bibitem{dislocation1} E. M. Chudnovsky and D. A. Garanin, preprint cond-mat/0105195
(2001).  

\bibitem{dislocation2} D. A. Garanin and E. M. Chudnovsky, preprint cond-mat/0105518
(2001).

\bibitem{zeldov} For details see E. Zeldov, D. Majer, M. Konczykowski, V. B. Geshkenbein, V. M. Vinokur, 
and H. Shtrikman, Nature {\bf 375}, 373 (1995); D. Majer, E. Zeldov, H. Shtrikman, 
and M. Konczykowski, in {\it Coherence in High Temperature Superconductors}, eds. 
G. Deutscher and A. Revcolevschi, World Scientific (Singapore, 1996) pp. 271-296.

\bibitem{Landau} L. D. Landau, Phys. Z. Sowjetunion {\bf 2}, 46 (1932).

\bibitem{Zener} C. Zener, Proc. R. Soc. London A {\bf 137}, 696 (1932).

\bibitem{Miyashita} S. Miyashita, J. Phys. Soc. Jpn. {\bf 64}, 3207 (1995).

\bibitem{dobrovitsky} V. V. Dobrovitsky and A. K. Zvezdin, Europhys. Lett. {\bf 38},
377 (1997).

\bibitem{leuenberger} M. N. Leuenberger and D. Loss, Phys. Rev. B {\bf 61}, 12200
(2000).

\bibitem{shift} Since we have plotted Fig. \ref{UnscaledRelaxation} on a
log scale, replotting the data in terms of
$\Delta_{N,i}$ would only result in a change of scale of the horizontal axis.

\end{references}
\end{document}